# Correlation Pseudogaps in HTSC

**Moshe Dayan**

**Department of Physics, Ben-Gurion University,**

**Beer-Sheva 84105, Israel.**

E-mail: mdayan@bgumail.bgu.ac.il

# Correlation Pseudogaps in HTSC

Moshe Dayan

Department of Physics, Ben-Gurion University,

Beer-Sheva 84105, Israel.

## Abstract

The electronic correlations in oxide HTSC in the normal state are analyzed by the generalized Hartree-Fock theory of Nambu and Gorkov. It yields two kinds of pseudogaps. The small pseudogap (SPG) is obtained as the Fock diagram, whereas the large pseudogap (LPG) is obtained as the sum of the Fock and the Hartree diagrams. We find real gaps for the undoped insulating materials, and complex pseudogaps for the doped and conducting materials. The latter result in the V-shaped experimental density of states.

Oxide HTSC are obtained by doping materials that are supposed to be metallic by band structure considerations, but are actually insulators. Some of these "mother" materials are antiferromagnetic (AF) insulators, but others are insulators without any magnetic characteristics. Such is for example $BaBiO_3$, which was proposed to have Bi atoms of alternating valence ($Bi^{+3}$ and $Bi^{+5}$), which makes the material a CDW insulator. The superconductivity of the BKBO family, which is derived from $BaBiO_3$, is comparable with the superconductivity of the LSCO family, which is derived from the AF $La_2CuO_4$. So are many features in the normal state. The BKBO family demonstrates that AF is not essential to superconductivity. Our view is that AF is not essential to establish high $T_c$ superconductivity, but that both characteristics stem from the same source. In both magnetic and non-magnetic HTSC, there are strong electron correlations, due to the strong nesting that exists in many materials of the perovskite structure and its 2-dimentional associates. Recently, the present author has shown that this nesting leads to divergences of electronic polarizations and of the dielectric function, at zero frequency [1]. Such irregularities usually trigger the breakdown of the symmetry of the electronic states.

Thus, we are dealing with a symmetry breakdown that is caused by electron correlations. This suggests the use of the Generalized Hartree-Fock (GHF) method of Nambu and Gorkov [2A, 2B], since this method was proved most successful in treating the transition to the superconductive symmetry, which is caused by electron correlations [2,3]. Here we use the Nambu's scheme to analyze the normal state of HTSC, and their mother insulating materials. The inherent differences from superconductivity imply suitable differences in the definition of certain functions, and in their treatment, but the main features are similar. In the problem of superconductivity, the off diagonal self-energy is the pairing self-energy. Here, the off diagonal self-energy is the gap in the undoped materials, and the pseudogap in the doped HTSC. Similarly, the field operator is defined differently. We assume that in the undoped materials most states by the Fermi surface are nested, and consequently, take part in the transition to the condensed phase. In the doped materials, however, only a part of the Fermi surface undergoes quasi-phase transition. The other part is renormalized in the usual manner. This suggests lengthy definition procedure, which is prohibited in the present letter because of its limited volume. A more detailed paper will be published elsewhere [4].

Here, we define the field operator as

$$\Psi_{ks} = \begin{pmatrix} c_{k,s} \\ c_{\bar{k},s} \end{pmatrix}, \text{ and its conjugate } \Psi_{ks}^+ = \begin{pmatrix} c_{k,s}^+ & c_{\bar{k},s}^+ \end{pmatrix}, \tag{1}$$

where $\bar{\mathbf{k}} = \mathbf{k} - (\mathbf{k}/|\mathbf{k}|) \cdot 2k_F$, $\mathbf{k}$ is the state wavevector, s the spin, and c and $c^+$ are the usual annihilation and creation operators. The propagator G is a 2x2 matrix, and by analogy with the scalar case, is given by

$$G_s(k,t) = -i < \Phi | T\{\Psi_{ks}(t)\Psi_{ks}^+(0)\} | \Phi >$$

$$= -i < \Phi | T \begin{pmatrix} c_{k,s}(t)c_{k,s}^+(0) & c_{k,s}(t)c_{\bar{k},s}^+(0) \\ c_{\bar{k},s}(t)c_{k,s}^+(0) & c_{\bar{k},s}(t)c_{\bar{k},s}^+(0) \end{pmatrix} | \Phi >, \tag{2}$$

where T is the usual time ordering operator, and $|\Phi>$ is the ground state of the interacting system. In perturbation field theory, the propagator is expanded as a Feynman-Dyson perturbation series, which is given by operators in the "interaction picture", acting on the ground state of the Hamiltonian without interactions. An essential notion of any theory of phase transitions is that one cannot obtain the condensed phase as a perturbation on the non-condensed phase. This is so because of the different symmetry of the ground states of the two phases. To get around this difficulty, one must define an unperturbed Hamiltonian whose ground state has the symmetry of the condensed phase. To obtain this quality Nambu added a one-body off-diagonal Hartree-Fock potential, which produces the right symmetry, and which was assumed to be determined self-consistently by the perturbation theory [2.3]. Here we write the Hamiltonian by means of the field operators $\Psi$, and the off-diagonal one-body potential $\Lambda_0$

$$H_0' = H_0 + \frac{1}{2}\sum_{k,s}\Lambda_0 \Psi_{k,s}^+ \tau_1 \Psi_{k,s}, \tag{3a}$$

$$H_i' = H_i - \frac{1}{2}\sum_{k,s}\Lambda_0 \Psi_{k,s}^+ \tau_1 \Psi_{k,s}, \tag{3b}$$

$$H_i = \frac{1}{8}\sum_{k,k',q,s,s'}\{V_{I,q}[\Psi_{k'-q,s'}^+ I\Psi_{k',s'}\Psi_{k+q,s}^+ I\Psi_{k,s}]$$

$$+ V_{1,q}[\Psi_{k'-q,s'}^+ \tau_1 \Psi_{k',s'}\Psi_{k+q,s}^+ \tau_1 \Psi_{k,s}]\} \tag{3c}$$

$$H = H_0 + H_i = H_0' + H_i' \tag{3d}$$

where $H_0$ is the regular unperturbed Hamiltonian, $\tau_1$ is the well-known Pauli matrix, and the V's are the interaction potentials. Notice the $\tau_1$ and the I interaction vertices in Eq. (3c). Also note that the numerical prefactors in Eqs. (3a-c) are different than usual, due to the fact that each interaction term ( with the c operators) repeats itself when the summations are carried out. An alternative notation would be to restore the numerical prefactors to the usual notation, but to perform the momentum summation only on states inside the Fermi surface. Note that, although the $\tau_1$ vertex in Eq. (3c) appears formally as transferring a momentum **q**, it actually transfers a momentum of **q+2k$_F$**. Note also that we dropped from the Hamiltonian the phononic part, and the electron-phonon interaction part, just to avoid lengthy writing and discussion. The el-phonon interaction will be taken into consideration when we apply perturbation theory. We also did not put in the Hamiltonian elastic scatterings by impurities. These scatterings will be analyzed in the last part of the paper.

The ground state of the unperturbed Hamiltonian of Eq. (3a) has the symmetry of the condensed state, and is given by

$$|\Phi_0> = \prod_{E_k<0,s}(u_k + v_k c^+_{\bar{k},s} c_{k,s})|0> = \prod_{E_k<0,s}|\Phi_{k,s}>, \tag{4}$$

where $|u_k|^2 + |v_k|^2 = 1$, $E_k$ are the eigenvalues of $H_0$, and $|0>$ is its ground state. The definition of the field operators as vectors, defines two kinds of products and consequently, two kinds of anti-commutation relations. Since each product is between a row vector and a column vector, one gets a scalar when the column vector is on the right side, and a matrix when it is on the left side. In the following relations the relative order is kept as predetermined regardless of the relative positions of $\Psi$ and $\Psi^+$ ( which suggests that $\Psi$ might also be a row vector, and $\Psi^+$ a column vector).

$$\{\Psi^+_{k,s}, \Psi_{k',s'}\} = 2\delta_{s,s'}\delta_{k,k'} . \tag{5a}$$

$$\{\Psi_{k,s}, \Psi^+_{k',s'}\} = \delta_{s,s'}[I\delta_{k,k'} + M\tau_1\delta_{k,\bar{k}'}], \tag{5b}$$

$$\{\Psi_{k,s}, \Psi_{k',s'}\} = \{\Psi^+_{k,s}, \Psi^+_{k',s'}\} = 0 , \tag{5c}$$

where the matrix M in Eq. (5b) is diagonal. The first of Eqs. (5) is scalar, the second is matrix, and the third may be both. The use of the Wick's theorem and the perturbation theory is justifiable as in the problem of superconductivity. The off diagonal part of Eq. (5b), is irrelevant, since we have used the notation in which k and k' are inside the Fermi surface, which implies that the delta function vanishes. Consequently, the perturbation theory is applicable with the replacement of the regular field operators by the vector field operators $\Psi$. The matrix propagator is given by

$$G(\omega,k) = \frac{I(\omega - \Sigma_I) + \tau_3 E_k + \tau_1 \Sigma_1}{(\omega - \Sigma_I)^2 - E_k^2 - \Sigma_1^2} \quad . \tag{6}$$

In Eq. (6) and hereafter, the spin indices are dropped for convenience. The self-energies $\Sigma_I$ and $\Sigma_1$ should be determined self-consistently.

We start with the off diagonal self-energy. In the undoped case it is the sum of two contributions, $\Sigma_1 = \Lambda = \Lambda_H + \Lambda_F$, where $\Lambda_H$ results from the known Hartree diagram, and $\Lambda_F$ from the Fock diagram. The Hartree contribution is given by

$$\Lambda_H = -iU_{1,0}(\omega = 0)\text{Tr}(\tau_1^2)\int\frac{d^3p d\nu}{(2\pi)^4}\frac{\Lambda \exp(i\nu\delta)}{\nu^2 - E_p^2 - \Lambda^2} \quad . \tag{7}$$

In Eq. (7) $U_{1,0}(\omega = 0)$ is the fully dressed static interaction that corresponds to $q = 2k_F$. It includes Coulomb interaction as well as el-phonon-el interaction. Note that in Eq. (7), and throughout the whole paper, the momentum integration is presented as 3-dimentional. This is only a formal presentation, and should not invalidate the analysis for 2-dimentional HTSC. The Fock integral is given by

$$\Lambda_F(\omega,k) = \frac{i}{2} \int \frac{d\nu d^3p}{(2\pi)^4} G_1(\nu,p) \sum_{j=0,1} \{\sum_\lambda g^2_{\lambda,jj} D_\lambda(\omega-\nu,k-p) + \exp(i\nu\delta) V_{jj}(k-p)\}, \quad (8)$$

where $D_\lambda$ is the propagator of the phonon of mode $\lambda$, g is the el-phonon matrix element, and the subscript j denotes the interaction vertex. Equations (7) and (8), and the relation $\Lambda = \Lambda_H + \Lambda_F$, are coupled integral equations. To solve them we assume $\Lambda_F < \Lambda_H$, and take the ratio $f = \Lambda_F / \Lambda_H < 1$ as a small parameter. The justification for this assumption is discussed later. With this assumption Eq. (7) includes only $\Lambda$ on both sides of the equality sign, and may be integrated. We change its momentum integration into integration with respect to the band energy. For simplicity, we assume that significant nesting prevails within the energy range $\pm E_m$, about the Fermi level, and take $N_0$ to be the band DOS within this range. We get

$$\Lambda^2 - 2\Lambda E_m \exp(-\frac{1-f}{2UN_0}) + 4E_m^2 = 0. \quad (9)$$

We assume U to be negative since the el-phonon-el attraction in these materials exceeds the Coulomb repulsion. This should eliminate one of the two solutions because it is too large to agree with any experiment. So that we are left with the other solution

$$\Lambda = 2E_m \exp(\frac{1-f}{2UN_0}). \quad (10)$$

Eq. (10) yields only real values for $\Lambda$. It is straightforward to show that real $\Lambda$ implies the existence of a real gap (and zero DOS within this gap). This is in accord with experiment in insulating undoped materials. For doped and conducting HTSC, experiments exhibit states in the pseudogaps, and also linear increase of the DOS on both sides of the Fermi level [5-7]. When an insolating "mother" material is doped to yield conducting HTSC, the Fermi level is shifted away from its original mid band value, and nesting is reduced. The doping also introduces disorder. The standard treatment of impure metals assumes elastic scatterings by randomly distributed impurity atoms. The Feynman diagram that depicts double scatterings from the same impurity atom yields complex self-energy even at zero energy [8]. Here we apply the same treatment to our condensed (or quasi-condensed) electron states. As in the regular problem, we assume that the same process results in a relaxation rate $\gamma$, and calculate it self consistently. The propagator of the doped HTSC is

$$G = \frac{1}{2\sqrt{[\nu + i\gamma S(\nu)]^2 - \Lambda^2}} \{(E+B)^{-1} - (E-B)^{-1}\}\{I[\nu + i\gamma S(\nu)] + \tau_3 E + \tau_1 \Lambda\}, \quad (11)$$

where $B = \sqrt{[\nu + i\gamma S(\nu)]^2 - \Lambda^2}$, and $S(\nu) = \nu/|\nu|$. Note that, apart from the different form of writing, Eq. (11) is different from Eq. (6) only by the introduction of the finite relaxation rate $\gamma$, and by the implicit assumption that $\Lambda$ is complex. The double elastic scattering process from the same impurity requires only momentum integration, and it is given by

$$\Lambda_i = N_i \sum_{j=0,1} \int \frac{d^3p}{2(2\pi)^3} W_{kp}^j W_{pk}^j \frac{I[\nu + i\gamma S(\nu)] + \tau_1 \Lambda}{\sqrt{[\nu + i\gamma S(\nu)]^2 - \Lambda^2}} [(E_p + B)^{-1} - (E_p - B)^{-1}]. \quad (12)$$

In Eq. (12), $N_i$ is the density of impurities, $W_{kp}^j$ is the Furrier transformed scattering potential, and the superscript j denotes the vertex. This disorder-induced matrix self-energy is complex even at zero energy. Here we present only the imaginary parts of its components, because the real parts make only small corrections to the self-energy. Integration of Eq. (12) yields for small $\nu$

$$\Lambda_{i0} = -iS(\nu)\frac{\pi N_i N_0 |\overline{W}_k|^2 \gamma}{\sqrt{\gamma^2 + \text{Re}\,\Lambda^2}} = -iS(\nu)\gamma \qquad (13)$$

$$|\overline{W}_k|^2 \cong \int dE \frac{\{\sum_{j=0,1} W_{kp}^j W_{pk}^j\}\sqrt{\gamma^2 + \text{Re}\,\Lambda^2}}{E^2 + \gamma^2 + \text{Re}\,\Lambda^2} \qquad (14)$$

Eq. (14) may be considered as an integral average of $\sum_{j=0,1} W_{kp}^j W_{pk}^j$ by a "broad delta-function", whose width is $\sqrt{\gamma^2 + \text{Re}\,\Lambda^2}$. In the present analysis we are interested in small energies, so that the subscript k may be eliminated from the averaged interaction. For the off diagonal component we get

$$\Lambda_{i1} = N_i N_0 |\overline{W}|^2 \frac{\Lambda}{B}\{\frac{4\,\text{Re}\,B}{E_m} - i\pi\}. \qquad (15)$$

The imaginary part coincides with $\text{Im}\,\Lambda$, and may be solved

$$\text{Im}\,\Lambda = -\frac{4\gamma\,\text{Re}\,\Lambda}{\pi[\gamma + \sqrt{\gamma^2 + \text{Re}\,\Lambda^2}]} \qquad (16)$$

When $\gamma \gg \text{Re}\,\Lambda$, we have $\text{Im}\,\Lambda \cong -\frac{2}{\pi}\text{Re}\,\Lambda$.

The DOS is given by

$$N(\nu) = \frac{-S(\nu)}{\pi} \text{Im} \int dE N_0(E) G(\nu, E), \qquad (17)$$

which yields for the impure case at low energy

$$N(\nu) = N_0 \{ \frac{\gamma}{(\text{Re}\Lambda^2 + \gamma^2)^{1/2}} + \frac{-\text{Im}\Lambda^2 |\nu|}{2(\text{Re}\Lambda^2 + \gamma^2)^{3/2}} \} \qquad (18)$$

Eq. (18) includes two terms. The first term is independent of energy. It gives the constant disorder-induced DOS at the Fermi level and throughout the gap. This constant DOS adds to the metallic DOS of the non-nested parts of the Fermi surface that are not represented in our Hamiltonian (but will be discussed shortly in [4]). The second term is linear in $|\nu|$. This is the well-known V-shaped DOS that has been obtained by many measurements [5-7]. Note that the sign of $\text{Im}\Lambda$, as given by Eqs. (16), results in the proper experimental increase of the DOS.

The off-diagonal self-energy is obtained from integrals that are depicted by two diagrams, the Hartree diagram, and the Fock diagram. The ratio $f = \Lambda_F(\omega = 0)/\Lambda_H$ was assumed to be small without justification. A full study of the two integrals is needed for the adequate evaluation of f. However, even without such an extensive analysis, we can speculate regarding the nature of these two components. More than a decade has passed since physicists became convinced that pseudogaps exist in HTSC, and are responsible for part of their odd features in the normal state [9-13]. There is a distinction between small pseudogap (SPG) and large pseudogap (LPG), although the latter has seldom been discussed in the literature [10,11]. The SPG has been associated with magnetic character, and in magnetic HTSC in the under doped limit, the SPG has been regarded as spin pseudogap, or as originated by interaction with spin fluctuations [9,13]. It is well known that only the Fock self-energy diagram differentiates between spin states, whereas the Hartree diagram sums on both spin states. Consequently, we speculate that $\Lambda_F$ should be the SPG. The LPG should be associated with the total pseudogap $\Lambda$. Our analysis however, treats $\Lambda_F$ as off-diagonal self-energy that should exist in all HTSC (including the non-magnetic ones). The magnetic properties of $\Lambda_F$ should show up naturally in materials were the interaction strength versus band characteristics favor magnetic ordering or magnetic fluctuations. This is in accord with our treatment that considers the superconductivity of the BKBO to be not different in essence from the superconductivity of the cupper-oxides.

One of the important observations is the large isotope shift of the SPG in several cupper-oxides [12]. This is in agreement with our results. Although one can show that the direct contribution to the integral of $\Lambda_F$ at phonon energies is very small, the integral depends almost linearly on the total gap $\Lambda$, as can be seen from Eq. (8). The latter is strongly dependent upon U (as evident from Eq. (10)), and therefore, upon the el-phonon interaction.